\documentclass[aps,pra,groupedaddress,twocolumn]{revtex4-2}

\usepackage{graphicx}% Include figure files
\usepackage{dcolumn}% Align table columns on decimal point
\usepackage{amsmath}
\usepackage{bm}% bold math
\begin{document}

%\preprint{APS/123-QED}

\title{All-optical atomic magnetometry using an elliptically polarized\\amplitude-modulated light wave}

\author{Anton Makarov$^{1,2}$}
\author{Katerina Kozlova$^{1,2}$}
\author{Denis Brazhnikov$^{1,2}$}
\email[Corresponding author: ]{x-kvant@mail.ru}
\author{Vladislav Vishnyakov$^{1}$}
\author{Andrey Goncharov$^{1,2,3}$}

\affiliation{$^1$Institute of Laser Physics SB RAS, 15B Lavrentyev Avenue, Novosibirsk 630090, Russia}
\affiliation{$^2$Novosibirsk State University, 1 Pirogov Street, Novosibirsk 630090, Russia}
\affiliation{$^3$Novosibirsk State Technical University, 20 Karl Marks Avenue, Novosibirsk 630073, Russia}

\date{\today}

\begin{abstract}
We study a resonant interaction of an elliptically polarized light wave with $^{87}$Rb vapor (D$_1$ line) exposed to a transverse magnetic field. A $5$$\times$$5$$\times$$5$~mm$^3$ glass vapor cell is used for the experiments. The wave intensity is modulated at the frequency $\Omega_m$. By scanning $\Omega_m$ near the Larmor frequency $\Omega_L$, a magnetic resonance (MR) can be observed as a change in the ellipticity parameter of the wave polarization. This method for observing MR allows to significantly improve the signal-to-noise ratio compared to a classical Bell-Bloom scheme using a circularly polarized wave. The sensitivity of the magnetic field sensor is estimated to be $\approx\,$$130$~fT/$\surd$Hz in a $2$~kHz bandwidth, confidently competing with widely used Faraday-rotation Bell-Bloom schemes. The results can be used to develop a miniature all-optical magnetic field sensor for medicine and geophysics.
\end{abstract}

\maketitle

\section{\label{sec:1}Introduction}

Atomic magnetometry plays an important role in contemporary fundamental and applied research. Optically pumped magnetometers (OPMs) are used, for example, in the search for Dark Matter \cite{Afach2021} and the neutron electric dipole moment \cite{Abel2020}. Fundamental investigations with OPMs are also carried out in biology \cite{Fabricant2021}. In medicine, OPM-based sensors are being developed for magnetoencephalography (MEG) \cite{Alem2023}, magnetocardiography (MCG) \cite{Yang2021}, and other types of diagnostics.

To date, a number of atomic magnetometry methods have been developed \cite{Aleksandrov2009, Fabricant2023}. One the most widely used methods, also known as the Bell-Bloom scheme \cite{Bell1961}, employs a single amplitude-modulated (AM) circularly polarized light wave for both pumping and probing the atoms. If a magnetic field ${\bf B}$ is present in the cell, the light-induced macroscopic magnetic moment ${\bf M}$ of the medium experiences the Larmore precession at the frequency $\Omega_L$$\,=\,$$\gamma$$|{\bf B}|$, where $\gamma$ is the gyromagnetic ratio, which is equal to $2\pi$$\times$$7$~Hz/nT for the $^{87}$Rb ground state. If the frequency of the AM matches the Larmor frequency, $\Omega_m$$\,=\,$$\Omega_L$, the amplitude of the light intensity modulation exiting the cell reaches its maximum and decreases as $\Omega_m$ shifts away from $\Omega_L$. Thus, using a lock-in amplifier (LIA), a magnetic resonance (MR) can be observed by scanning $\Omega_m$ near $\Omega_L$.

Bell-Bloom magnetometers (BBMs) perform absolute measurements without the need for calibration of Helmholtz coils. They have a large dynamic range, allowing to perform measurements with subpicotesla sensitivity under the Earth's magnetic field conditions with a bandwidth of $\sim\,1$~kHz \cite{Zhang2019}, which can be extended up to $10$~kHz \cite{Martinez2012}. BBMs represent an all-optical technique, so there is no a cross-talk problem when using an array of closely placed magnetic sensors, unlike rf OPMs \cite{Aleksandrov2009,Martinez2010}.

\begin{figure*}[!t]
\centering
\includegraphics[width=\linewidth]{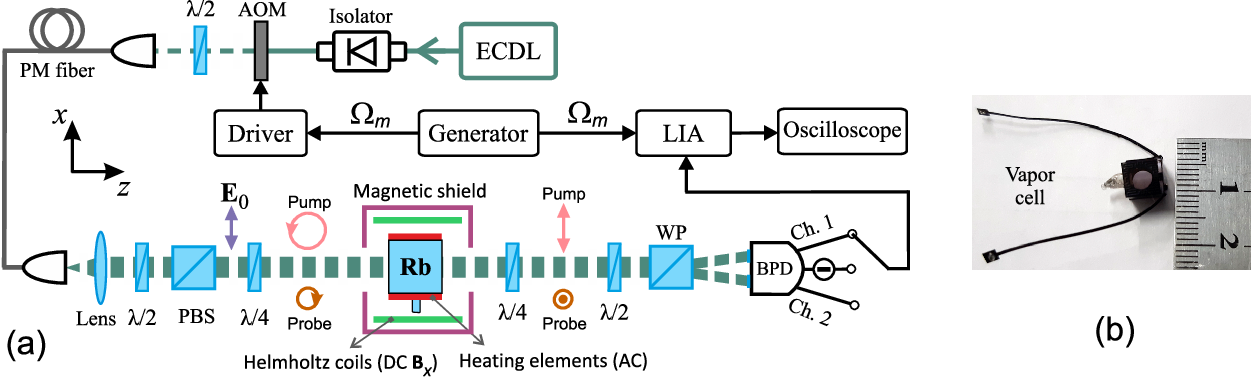}
\caption{(a) Experimental setup: ECDL, external-cavity diode laser; AOM, acousto-optic modulator; $\lambda$/2, $\lambda$/4, half-wave and quarter-wave plates, respectively; PM fiber, polarization maintaining fiber; PBS, polarizing beam splitter; WP, Wollaston prism; BPD, balanced photodetector; LIA, lock-in amplifier. (b) A rubidium vapor cell with heating elements.}
\label{fig:1}
\end{figure*}

Since it was first proposed \cite{Bell1961}, the Bell-Bloom scheme has undergone multiple modifications. In particular, AM of the light beam can be replaced by frequency modulation (FM) \cite{Martinez2010,Grujic2013} or polarization modulation \cite{Grujic2013,BenKish2010,Petrenko2021}. A linearly polarized pump wave can be utilized instead of a circularly polarized one \cite{Acosta2006,Pustelny2008,Wu2015}. BBMs are typically used for scalar measurements, although schemes for vector (three-axis) measurements have been proposed as well \cite{Patton2014,Huang2015,Grewal2020}. A significant improvement in the sensitivity of BBMs has been achieved by using a two-beam configuration with a modulated pump wave and a cw probe wave (here referred to as a pump-probe configuration). Nowadays, there is the demand for miniaturization of quantum sensors. In a pump-probe configuration, a magnetic sensor head can be miniaturized by using a single light beam \cite{Petrenko2021,Gerginov2020,Li2021}.

In commonly used pump-probe configurations, a linearly polarized probe wave is detuned far from the center of the absorption line (the rare exceptions are particular schemes with polarization modulation \cite{Petrenko2021} or amplitude modulation \cite{Li2021}, where an electro-optic modulator is required). In such a case, the probe wave experiences a very low absorption in the vapor cell, while its linear polarization is rotated due to the Faraday effect. An angle of this rotation provides the information about the external magnetic field. This effect is the result of the circular birefringence induced by a circularly polarized pump wave. OPMs based on this principle can demonstrate a signal-to-noise ratio of the order of $10^4$$-$$10^5$ in a $1$~Hz bandwidth \cite{Martinez2010}. However, large frequency detuning of the probe wave can introduce measurement errors due to optical frequency drifts over time. This issue forces researchers to develop methods to stabilize optical frequencies far from resonance with the medium \cite{Hu2017}.

In this work, we investigate a Bell-Bloom-like scheme with an elliptically polarized single light beam. It is in resonance with the $^{87}$Rb D$_1$ line ($\lambda$$\,\approx\,$$795$~nm). This beam can be decomposed into a superposition of two waves, the pump and probe waves, with counterrotating circular polarizations ($\sigma^+$ and $\sigma^-$). In contrast to schemes based on Faraday rotation, in our case the main contribution to the MR is caused by a circular dichroism of the medium. This phenomenon means that the pump and probe waves experience different absorption strengths. The latter leads to a change in the parameter of ellipticity of the light field polarization in the medium when $\Omega_m$ is being scanned around $\Omega_L$. This effect can be observed with the help of a polarimeter and gives information about the external magnetic field.

\section{\label{sec:2}Experiments}

\subsection{\label{sec:21}Experimental setup}

The experimental setup is shown in Fig.~\ref{fig:1}(a). An external-cavity diode laser (ECDL) in the Littrow configuration is used, which radiation linewidth was less than $1$~MHz. The wavelength was continuously varied by means of a diffraction grating equipped with a piezoelectric actuator. The optical frequency was controlled by a WS7 wavelength meter (Angstrom Co. Ltd.) with a resolution of $500$~kHz. The optical frequency was tuned to the center of the dipole transition $F_g$$\,=\,$$2$$\,\to\,$$F_e$$\,=\,$$1$. Due to the high pressure of the nitrogen buffer gas in the cell ($280$~Torr), the neighboring optical transitions $F_g$$\,=\,$$2$$\,\to\,$$F_e$$\,=\,$$1$ and $F_g$$\,=\,$$2$$\,\to\,$$F_e$$\,=\,$$2$ were not spectrally resolved and formed a single absorption contour.

\begin{figure}[!b]
\centering
\includegraphics[width=0.9\linewidth]{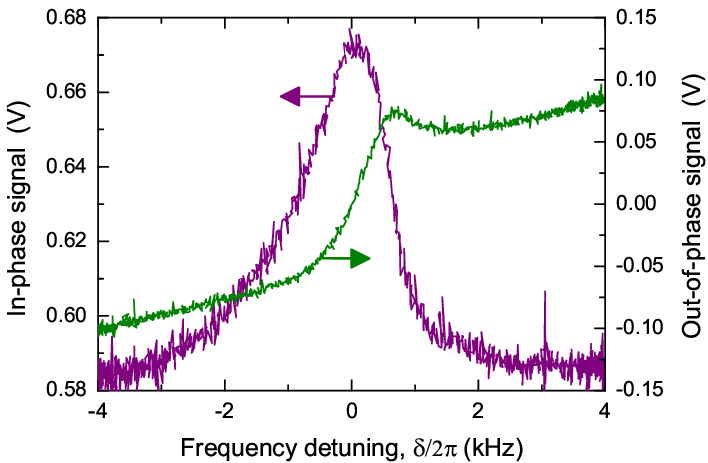}
\caption{\label{fig:2} A magnetic resonance in the classical Bell-Bloom scheme: in-phase (purple) and out-of-phase (green) signals of LIA. $T_{cell}$$\,=\,$$85^\circ$C, $P$$\,=\,$$800$~$\mu$W, $\epsilon$$\,=\,$$45^\circ$. Time constant of the LIA was $100$~$\mu$s, which corresponds to a $\approx\,$$1.6$~kHz bandwidth.}
\end{figure}

\begin{figure*}[!t]
\includegraphics[width=1\linewidth]{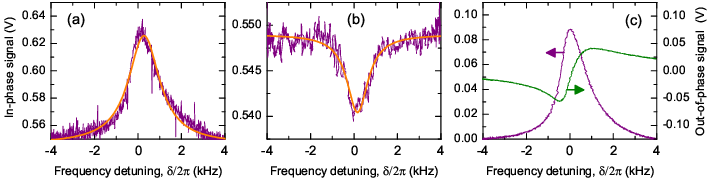}
\caption{\label{fig:3}A magnetic resonance in the proposed pump-probe scheme. The in-phase signals in Ch.1 (a) and Ch.2 (b) are fitted by a Lorentz shape. (c) The in-phase (purple) and out-of-phase (green) signals at the differential channel. $T_{cell}$$\,=\,$$85^\circ$C, $P$$\,=\,$$1.5$~mW, $\epsilon$$\,=\,$$20^\circ$. Time constant of the LIA was $100$~$\mu$s.}
\end{figure*}

The laser radiation passed through a Faraday optical isolator to eliminate the effect of spurious back reflections. An acousto-optic modulator (AOM) with an operating frequency $80$~MHz was used to perform AM of the light in the first diffraction order at a frequency $\Omega_m$. The beam then entered a polarization maintaining optical fiber. At the output of the fiber, the beam was expanded to a diameter of $\approx\,5$~mm (FWHM) using an adjustable fiber collimator and an additional lens. A combination of a half-wave plate ($\lambda/2$) and a polarizing beam splitter allowed continuous adjustment of the optical power at the entrance of the vapor cell.

A quarter-wave plate ($\lambda/4$) in front of the cell made it possible to control an ellipticity parameter ($\epsilon$) of the light field: $-\pi/4\,$$\leq\,$$\epsilon$$\,\leq$$\,\pi/4$, where $\epsilon$$\,=\,$$0$ means linear polarization, while $\epsilon$$\,=\,$$\pm\pi/4$ means circular polarization. In our case, it is convenient to expand the elliptically polarized beam into a superposition of two waves with opposite circular polarizations. For instance, $\epsilon$$\,=\,$$\pi/4$ for the pump wave (${\bf E}_c$) and $\epsilon$$\,=\,$$-\pi/4$ for the probe wave (${\bf E}_p$). The ratio of the pump and probe wave strengths is related to the ellipticity parameter as follows: $|{\bf E}_c|$$/$$|{\bf E}_p|$$\,=\,$${\rm ctg}(\epsilon-\pi/4)$. After passing through the cell and the second $\lambda/4$ plate, both light waves acquire linear polarizations orthogonal to each other. In the cell, different absorption coefficients for the probe wave and for the pump wave lead to a change in the ellipticity of the polarization compared to its initial value. To measure this, the beam was incident on a polarimeter consisting of a $\lambda/2$ plate, a Wollaston prism (WP) and a balanced photodetector (BPD).

A $5\times5\times5$~mm$^3$ cubic borosilicate glass vapor cell was used in the experiments. The cell was heated by thin resistive elements mounted on its side and front faces [Fig.~\ref{fig:1}(b)]. Heating was realized with ac current at $100$~kHz, which had no visible effect on the resonance parameters. The cell was enclosed in a three-layer
magnetic shield, which provided a residual field at the cell site of $\lesssim\,$$10$~nT (a shielding factor of the order of $10^4$).

To observe MR, $\Omega_m$ was scanned slowly ($5$~Hz) around the Larmor frequency $\Omega_L$, which was defined by a dc magnetic field applied to the vapor cell along the $x$ axis using a pair of Helmholtz coils. Scanning $\Omega_m$ near $\Omega_L$ led to a sharp change in the amplitude of the light intensity modulation at the exit of the cell due to MR. A lock-in amplifier SR865A (Stanford Research Systems) was employed to observe this effect. In-phase or out-of-phase (quadrature) signals of LIA were studied at oscilloscope.

\subsection{\label{sec:22}Results of measurements}

First, we examined an original Bell-Bloom scheme with a circularly polarized light. In this case, a polarimeter was not used. Fig.\ref{fig:2} shows in-phase (purple curve) and out-of-phase (green) signals as the functions of the frequency detuning $\delta$$\,=\,$$\Omega_m$$-$$\Omega_L$. In all experiments, the resonance's linewidth was of the order of $1$~kHz. Next, a pump-probe configuration was studied. As mentioned above, an elliptically polarized beam can be treated as a superposition of two waves with counterrotating circular polarizations. When there was no $\lambda/2$ plate in front of WP, each of the two waves was directed to a particular channel of the BPD: the pump wave was sent to a channel 1 (Ch.1), while the probe wave was sent to a channel 2 (Ch.2). A differential channel (Diff.) of the BPD provided a difference of the signals from Ch.1 and Ch.2. For better suppression of the light intensity noise at the Diff. channel, an additional $\lambda/2$ plate was used in front of the WP. The magnetic resonances observed at three channels of the BPD are shown in Fig.~\ref{fig:3}. It is clearly seen that the light intensity noise at the Diff. channel [Fig.~\ref{fig:3}(c)] is significantly reduced compared to each channel separately [Fig.~\ref{fig:3}(a,b)].

In this Letter, we do not discuss a theory of the resonance's lineshape in details. If the probe wave intensity is much smaller than the pump wave intensity, i.e. $\epsilon$ is relatively far from zero, then we can neglect the influence of the probe wave on the magnetic moment orientation. In this case, the main physical reasons for the signals observed in Ch.1 and Ch.2 are the same as for the original Bell-Bloom scheme\cite{Bell1961}. Let us just explain why the Lorentz profiles observed in Ch.1 and Ch.2 [Fig.~\ref{fig:3}(a) and Fig.~\ref{fig:3}(b), respectively] have different signs. For convenience, an MR that looks like Fig.~\ref{fig:3}(a) will be called a magnetic resonance of transparency (MRT), while the resonance in Fig.~\ref{fig:3}(b) will be referred to as a magnetic resonance of absorption (MRA).

A sign of resonance is the result of light-induced dichroism. A similar reason was responsible for a sign of the zero-field level-crossing resonances previously studied in pump-probe configurations \cite{Brazhnikov2020,Brazhnikov2022}. For simplicity, we consider a square-wave AM of the beam. It is also assumed that a commonly used condition $\Omega_m$$T_{1,2}$$\gg$$1$ is satisfied, where $T_1$ and $T_2$ are longitudinal and transverse relaxation times, respectively. An absolute value of the magnetic moment is assumed to be normalized to one ($|{\bf M}|$$\,=\,$$1$). The absorption of the pump wave in the vapor cell is proportional to $(1$$-$$M_z)$ \cite{Bell1961}, where $M_z$ is the $z$-projection of ${\bf M}$. The pump wave is assumed to be intense enough to polarize most atoms along the $z$-axis, i.e., $M_z$$\,\approx\,$$1$ when this wave is turned on. Therefore, the polarized atoms weakly scatter the pump wave and the medium becomes transparent for this wave. During the period between two pump pulses (a dark period), ${\bf M}$ undergoes free precession around the external magnetic field vector ${\bf B}_x$ at the frequency $\Omega_L$. In resonance, $\Omega_m$$=$$\Omega_L$, the momentum completes a full rotation until the pump field is turned back on and the medium remains transparent to the pump wave. Out of resonance, $\Omega_m$$\,\ne\,$$\Omega_L$, ${\bf M}$ does not complete a full rotation during the dark period. This leads to an increased pump-wave absorption in the cell, because the photons are scattered again to polarize the atoms along the $z$-axis. This is how the MRT type of resonance is formed [Fig.~\ref{fig:3}(a)].

For the probe wave, the light-atom interaction behaves in the opposite way, because the probe wave absorption is proportional to $(1$$+$$M_z)$. It means that at $M_z$$\,\approx\,$$1$ this wave experiences strong absorption in the cell. The absorption is reduced, if $\Omega_m$$\,\ne\,$$\Omega_L$, because in this case $M_z$$\,<\,$$1$. Therefore, scanning $\Omega_m$ near $\Omega_L$ leads to the observation of MRA instead of MRT in Ch.2 of the BPD [Fig.~\ref{fig:3}(b)]. Obviously, if the intensities of two waves are almost the same, they will compete for polarizing the atoms in different directions (along or against the $z$-axis). The latter will lead to the destruction of the MR in both channels. There is, therefore, an optimal ellipticity of the light polarization for achieving the highest signal-to-noise ratio at the Diff. channel ($\epsilon_{opt}$$\,\approx\,20^\circ$ in our experiments).

\begin{figure}
\centering
\includegraphics[width=1\linewidth]{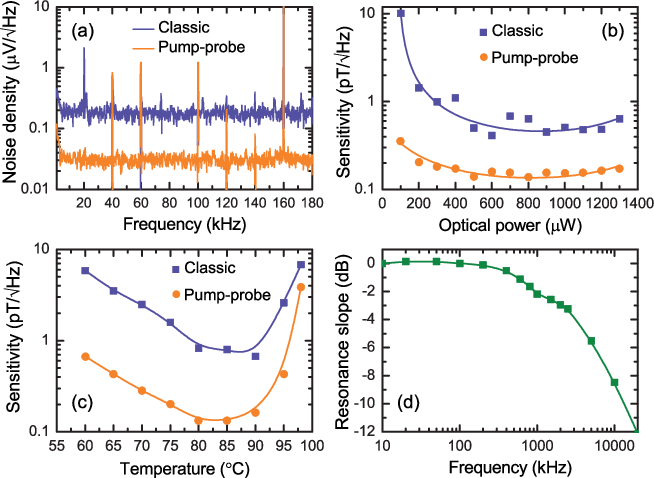}
\caption{(a) Noise voltage spectral density of signal from a photodetector in the classical scheme (violet, Ch.1, $P$$\,=\,$$800$~$\mu$W, $\epsilon$$\,=\,$$45^\circ$) and in the pump-probe scheme (orange, Diff. channel, $P$$\,=\,$$1.5$~$\mu$W, $\epsilon$$\,=\,$$20^\circ$). $T_{cell}$$\,=\,$$85^\circ$C. (b) and (c) reflect the sensitivity of magnetic field measurements versus optical power and temperature of the atoms, respectively. Here, in the classical scheme: $\epsilon$$\,=\,$$45^\circ$; $T_{cell}$$\,=\,$$90^\circ$C in (b) and $P$$\,=\,$$600$~$\mu$W in (c). In the pump-probe scheme: $\epsilon$$\,=\,$$20^\circ$, $T_{cell}$$\,=\,$$85^\circ$C in (b) and $P$$\,=\,$$600$~$\mu$W in (c). (d) Frequency response of the sensor in the pump-probe scheme at $T_{cell}$$\,=\,$$85^\circ$C, $P$$\,=\,$$1$~mW, $\epsilon$$\,=\,$$25^\circ$. Time constant of the LIA was $10$~$\mu$s. Solid curves in (b), (c) and (d) are just a guide for the reader's eye.}
\label{fig:4}
\end{figure}

Fig.~\ref{fig:4}(a) shows the light intensity noise measurements in the proposed scheme (orange data) compared to the classical scheme (violet). The data were acquired with the help of a spectrum analyzer SR1 (Stanford Research Systems). Since the amplitudes of the MR were the same in both schemes, it follows directly from the figure that the signal-to-noise ratio was improved by almost an order of magnitude. In Fig.~\ref{fig:4}(a), a sharp peak at $160$~kHz corresponds to the AM frequency of the light field. Other spurious peaks are associated with technical noise.

The sensitivity of a magnetometer, as the smallest detectable change in the magnetic field, can be estimated using the following expression \cite{Pustelny2008,Aleksandrov2009}: $\delta B$$\,\approx\,$$\Delta$$/$$(\gamma\times {\rm SNR})$. Here, $\Delta$ is a half width at half maximum (HWHM) of the MR, SNR is a signal-to-noise ratio in a given bandwidth. Fig.~\ref{fig:4}(b) reflects the behavior of $\delta B$ in a $1$~Hz bandwidth as the light power ($P$) is changed. The figure reveals an optimal value of $P_{opt}$$\,\approx\,$$800$~$\mu$W, which provides the highest sensitivity (the lowest $\delta B$). Another plot in Fig.~\ref{fig:4}(c) gives information of an optimal temperature, laying in the range of $80$$-$$85^\circ$C. Such dependencies are typical for Bell-Bloom-like schemes  \cite{Martinez2012,Petrenko2021}.

We estimated a sensitivity of the sensor to be about $130$~fT/$\surd$Hz under current experimental conditions. Investigating different sources of technical noise in our experiments (e.g., current noise of the Helmholtz coils, noise of the heating elements of the cell, etc.) and eliminating them will further improve the sensitivity. The maximum achievable sensitivity ($\delta B_{sn}$) can be obtained in the photon shot-noise limit. In this case, ${\rm SNR}$$\,=\,$$\sqrt{N_{ph}}$, with $N_{ph}$ being a photon flux incident on a photodetector. In the regime of Fig.~\ref{fig:3}, each channel of the BPD receives $P$$\,\approx\,$$60$~$\mu$W. The corresponding SNR equals $\approx\,$$1.5\times10^7$, leading to $\delta B_{sn}$$\,\approx\,$$10$~fT/$\surd$Hz. A sensor bandwidth, which characterizes a speed of measurements, is determined by a linewidth of MR. Fig.~\ref{fig:4}(d) shows the frequency response of the resonance observed in the quadrature channel of the LIA. As seen, the $-3$~dB bandwidth is $\approx\,$$2$~kHz.

\section{Conclusions}

In this work, we proposed a simple and robust technique based on an elliptically polarized amplitude modulated single light beam resonant to the D$_1$ line in alkali-metal atoms. The current sensor sensitivity is superior to those obtained for other compact sensors based on a classical Bell-Bloom scheme \cite{Martinez2010,Huang2015,Wu2015} and confidently competes with results obtained with the Faraday-rotation Bell-Bloom schemes \cite{Gerginov2020,Li2021,Pustelny2008}. The achieved sensitivity and the bandwidth allow the proposed scheme to be applied to MCG. Further improvement of the SNR is feasible, which will make it possible to apply the scheme to MEG as well.

The proposed scheme is easier to implement than many other Bell-Bloom-like schemes. In particular, there is no need for a far detuned probe light beam as in the commonly used Faraday-rotation schemes. Thus, the optical frequency of the beam in our scheme can be stabilized at a center of the absorption line of the atoms, avoiding its drift over time. The AM of the radiation, implemented in our experiments with an AOM, can be replaced by frequency modulation (FM) \cite{Martinez2010}. This would not require an external modulator, since FM can be achieved by directly modulating the pump current of the diode laser. This approach will allow miniaturization of the magnetometer as well as a reduction in its power consumption. Since we employ a resonant light-atom interaction, there is no need for the use of a bandpass filter behind the vapor cell in contrast to some other single-beam pump-probe schemes \cite{Gerginov2020}. Finally, the use of modern nanophotonic devices pays the way to an extreme miniaturization of a sensor head down to $1$~cm$^3$. For example, a ``spin selector'' \cite{Sebbag2021} could replace all the optical elements we used after the vapor cell, including a balanced photodetector.

\begin{acknowledgments}
The work has been supported by Russian Science Foundation (Grant no. 23-12-00195).
\end{acknowledgments}

\end{document}